"Optical Study of the Stripe-Ordered State"
Setsuko Tajima[1)] and Shin-ichi Uchida[2)]
[1)]Department of Physics, Osaka University, Osaka 560-0043, Japan
[2)]Department of Physics, The University of Tokyo, Toyko113-0033 , Japan

Abstract

The effects of the stripe order on the optical spectra of La-based cuprates are reviewed. The main effect on the high $T_c$ superconducting cuprates is to rapidly reduce the Josephson plasma frequency in the $c$-axis spectrum as a consequence of weakening of the Josephson coupling between $CuO_2$ layers. This points toward a two dimensional (2D) superconductivity in the stripe phase, although it is difficult to realize a 2D superconductivity in real materials. We also discuss the experimental results suggesting the presence of stripe effect in other cuprates even if they do not show the static stripe phase. Compared to the $c$-axis spectra, the in-plane spectra are not so dramatically affected by the stripe order, showing a weak gap-like feature and reducing the condensate spectral weight.

Corresponding author:
Setsuko Tajima
Dept. of Physics, Osaka University
Machikaneyama 1-1, Toyonaka, Osaka 560-0043, Japan
Tel/fax: +81-6-6850-5755
e-mail: tajima@phys.sci.osaka-u.ac.jp

Content

1.Introduction

The spin and charge stripe order is commonly observed in strongly correlated electronic systems such as $(La,Ba)_2CuO_4$ [1,2], $(La,Nd,Sr)_2CuO_4$ [3-5], $(La,Sr)_2NiO_4$ [6-8], $(La,Sr)_2CoO_4$ [9], and $(La,Sr)_2MnO_4$ [10-12]. Here the strong electron-electron interaction is a dominant source of the charge segregation accompanied by the spin order. In the Ni-, Co- and Mn-oxides, the doped carriers are completely localized in the stripe order. By contrast, the stripe ordered phase in the cuprates remains conducting or even superconducting. This unique property of the cuprates raises an important question whether the stripe order (or its fluctuation) plays a crucial role in the superconductivity pairing or acts only as a competitor.

To address this problem, it is important to investigate the superconducting state of the stripe ordered phase. We overview in this article the optical spectra of the stripe materials which revealed a unique electronic state in the superconducting stripe phase. Before discussing the stripe effects, it is necessary to know first the characteristic optical properties of the cuprates, in particular, the $c$-axis optical spectrum. Thus, we begin with a brief introduction about the unique feature in the $c$-axis optical spectra of the superconducting cuprates. Then, we see how the stripe order affects the c-axis spectrum of the prototypical stripe systems, $La_{2-x}Ba_xCuO_4$ and $La_{2-x-y}Nd_ySr_xCuO_4$, and how the spectrum vary with increase of the hole doping level in the latter system. Based on these observations we consider if the stripe effect exists in other cuprate systems which do not show apparent stripe order. Finally, we review the effect of the stripe order on the in-plane spectrum.

1.1. Josephson plasma in the $c$-axis optical spectra

It is well known that the high $T_c$ superconducting cuprates (HTSC) are a highly anisotropic (quasi two-dimensional) system, which is evidenced by the semiconductor-like temperature dependence of $c$-axis resistivity and/or the insulator-like optical spectra for $E//c$ [13]. The $c$-axis optical spectra of HTSC, except for $YBa_2Cu_3O_y$ (Y123) and $YBa_2Cu_4O_8$ (Y124), are dominated by the optical-mode phonon peaks with little component of free carriers. This non-Drude like optical spectrum in the normal state originates from an extremely small plasma frequency as well as a large carrier scattering rate in the $c$-direction, suggestive of an incoherent charge dynamics along the $c$-axis [14].

When the system goes into a superconducting state, a sharp plasma edge is suddenly created in the c-axis reflectivity spectrum, as is shown in Fig. 1 [15,16]. The edge frequency or the screened plasma frequency $\omega_p'$ of La214 is lower than the lowest optical phonon frequency in the far-infrared region, and is also low enough to excite appreciable number of quasiparticles across the superconducting gap within the $CuO_2$ layer. As a consequence, the plasma edge appearing below $T_c$ is sharp without suffering strong damping. The manifestation of a sharp plasma edge indicates a recovery of the *c*-axis phase coherence, giving a support for the picture that Cooper pairs are formed within the $CuO_2$ layer and the layers are weakly coupled via the pair Josephson tunneling A schematic picture is illustrated in Fig.2. We call the plasma mode in such a condition as Josephson plasma mode [17-19].

In contrast to a dramatic change in the reflectivity spectrum, the conductivity spectrum shows only a suppression of conductivity below the gap energy without any sharp feature at the plasma edge energy. Figure 3 displays typical c-axis optical conductivity spectra for La214. The lost spectral weight below $T_c$ is condensed into a delta function at $\omega=0$ [20].

There are three ways to estimate a condensate weight at $\omega=0$ or a superconducting plasma frequency $\omega_{ps}$ in the *c*-direction. The first one is to estimate a spectral weight of a delta function from the missing area of $\sigma_1(\omega)$, based on the Ferrell-Glover-Tinkham sum rule[21].

$$(2\pi/\lambda_c)^2 = \omega_{ps}^2 = 8\int_{0^+}^{\infty}[\sigma_1^N(\omega)-\sigma_1^S(\omega)]d\omega, \quad (1)$$

if the kinetic energy contribution is negligibly small [22]. The second is to determine $\omega_{ps}$ directly from a reflectivity spectrum by fitting. The third is to estimate $\omega_{ps}$ from the low frequency $\sigma_2(\omega)$ or $\varepsilon_1(\omega)$ through the low-$\omega$ asymptotic relation $\varepsilon_1(\omega) \sim -\omega_{ps}^2/\omega^2$. The effect of stripe order on the superconductivity condensate can be examined by studying these quantities.

1.2. Transverse Josephson plasma in the modulated coupling systems

The Josephson-coupled-layer model well describes the *c*-axis charge dynamics of HTSC below $T_c$. Extending this simple model, we can expect an additional structure in the *c*-axis spectra. The Josephson coupling strength between the $CuO_2$ layers is strongly dependent on the nature of blocking layers. Therefore, when there are two or three different blocking layers per unit cell as illustrated in Fig.2, the interlayer coupling strength is not uniform but is periodically modulated along the *c*-axis. This modulation gives a transverse Josephson plasma mode which appears as a broad absorption peak in the far-infrared region below $T_c$ [23]. The mechanism is similar to the appearance of optical phonon mode from an acoustic phonon branch as a result of Brillouin zone folding.

A typical example of the transverse Josephson plasma is seen in single layer T*-214 compounds which are composed of a network of $CuO_5$ pyramids sandwitched by different blocking layers [24-27]. Figure 4 shows an example of the spectra for $SmLa_{0.85}Sr_{0.15}CuO_4$ and $Nd_{1.4}Sr_{0.4}Ce_{0.2}CuO_4$. A weak peak develops below $T_c$ just above the plasma edge frequency. The double layer HTSC which contain two $CuO_2$ layers in a unit cell are also good candidates for the modulated coupling system with different intra- and inter-bilayer coupling strengths. A similar or more pronounced peak due to the transverse Josephson plasma has been observed in most of the double layer HTSC such as Y123 [28], Y124 [29], $Pb_2Sr_2(Y/Ca)Cu_3O_8$ [30] and Bi2212 [31].

2. *c*-Axis Optical Spectra in the Stripe Order Phase
2.1 Interlayer dephasing in the stripe ordered state

To explore the optical response of the stripe ordered phase, the $La_{2-x-y}Nd_ySr_xCuO_4$ (La/Nd214) system is more appropriate than $La_{2-x}Ba_xCuO_4$ (LBCO). In LBCO the static stripe order is formed only near $x$ =1/8, whereas the stripe order in La/Nd214 is observed over a much wider range of $x$, and the interplay between superconductivity and stripe order can be studied by changing the Nd ($y$) content.

Figure 5 illustrates the phase diagram of $La_{1.85-y}Nd_ySr_{0.15}CuO_4$ [32, 33]. The compound exhibits a phase transition from the low-temperature-orthorhombic (LTO) to the low-temperature-tetragonal (LTT) structure (at $T_d$ = 75 K for y=0.4), which is signaled by a resistivity jump [32-34]. The spin/charge stripe order is presumably pinned by the LTT lattice distortions and it becomes almost static. Note that even in the LTT phase, superconductivity survives with a rather high $T_c$.

To see how the static stripe order is crucial for suppressing Josephson plasma frequency, we investigate the Nd content dependence of reflectivity spectra for a fixed Sr content. Figure 6 shows the 8K reflectivity spectra of $La_{1.85-y}Nd_ySr_{0.15}CuO_4$ for various $y$'s [35]. As Nd content increases to 0.12, the Josephson plasma edge rapidly shifts to lower frequency. When y exceeds 0.12, the plasma edge suddenly disappears. Concomitantly, the structural phase transition from LTO to LTT is observed only above $y$=0.12. This coincidence of the critical y value strongly suggests that the LTT phase that stabilizes the static stripe order radically weakens the interlayer Josephson coupling. The results are summarized in Fig. 7 which is a clear demonstration of the interlayer dephasing by the stripe order.

2.2 Static stripe phase in $La_{2-x}Ba_xCuO_4$ and Nd-doped $La_{2-x}Sr_xCuO_4$ with $x$ =1/8

No indication of a Josephson plasma edge is seen in the stripe ordered phases of $La_{2-x}Ba_xCuO_4$ [36] and $La_{1.6-x}Nd_{0.4}Sr_xCuO_4$ with $x$ equal to 1/8. It may not be surprising as the bulk (3D) $T_c$ is too low (below 4 K) to be accessed by an ordinary experimental setup. However, the transport measurement demonstrates that the superconducting order, though short ranged, develops within a layer at temperature ($T_{BKT} \sim$ 16K) well above the bulk $T_c$, where $T_{BKT}$ is a Berezinskii-Kosterlitz-Thouless transition temperature [37]. The reported in-plane spectrum in this state below $T_{BKT}$ seems to show dramatic Drude narrowing, instead of showing condensation to the $\omega$=0 delta function [36]. Then, the absence of a Josephson plasma edge in the c-axis spectrum below $T_{BKT}$ may give evidence for the loss of interlayer phase coherence. In the case of Nd-free La214 ($x$ = 0.12) with $T_c$ = 28 K, a sharp plasma edge is observed at $\sim$25 cm$^{-1}$ at $T$ = 8 K. Once Nd is incorporated, the reflectivity edge radically shifts to lower energy, say, down to 7 cm$^{-1}$ for y=0.2 ($T_c\sim$10K) [38].

The stripe order is most stable in these compounds. At this hole concentration both spin and charge order are almost commensurate with the underlying lattice – the period of spin-stripes (charge-stripes) is close to 8 (4) lattice constant of the $CuO_2$ layer, and hence matches the LTT lattice modulations. In this case the stripes are pinned by the lattice potential, and the stripe order is static with a fairly long range spatial correlation Therefore, the stripe order realized in these compounds would be nearly an ideal stripe order for which the theoretical models propose the interlayer decoupling [39, 40]. As illustrated in Fig.8, in the ideal stripe order the charge stripes on one layer run along the direction perpendicular to those in the neighboring layers, so the stripe-LTT structure is a stack of two 'orthogonal' layers, the Josephson coupling between which exactly cancels. Moreover, the parallel stripes in the second neighbor layers are staggered – shifted by half a period in order to minimize the Coulomb interactions, resulting in a further doubling of the number of layers per unit cell. In this stripe structure, the Josephson coupling between 2nd and 3rd neighboring layers also cancel, so the non-vanishing coupling occurs between those 4 layers apart which would be very weak.

2.3 Stripe phase in Nd/Eu-doped $La_{2-x}Sr_xCuO_4$ near the optimum doping ($x \sim$ 0.15)

In Fig. 7 are also plotted the values of the superconducting $T_c$ as a function of y for $La_{1.85-y}Nd_ySr_{0.15}CuO_4$. One sees that, even in the static stripe phase ($y$ >0.12) where the Josephson plasma frequency is strongly reduced, the superconductivity survives, keeping a relatively high $T_c$ values ($T_c$ = 12 K for $y$ =0.4).A problem is that it is hard to judge whether the Josephson plasma really disappears or not because the measurable frequency range is limited down to $\sim$20 cm$^{-1}$ in the ordinary experimental configuration.

Recently Tanaka *et al.* have extended a frequency range down to 0.15 THz ($\sim$5 cm$^{-1}$) by using a THz time-domain spectroscopy (THz-TDS) [41]. They could find a Josephson plasma edge with much shallower dip in the sub-THz region for the sample with $T_c$ higher than 10 K. In Fig. 9 are shown the $c$-axis reflectivity spectra of $La_{1.84-y}Nd_ySr_{0.16}CuO_4$ for $y$ =0.3 and 0.4 [42]. For y = 0.3, a plasma edge develops at around 0.8 THz (~26 cm$^{-1}$) below $T_c$ (=20 K). For $y$ =0.4 ($T_c$ =12 K), only a weak increase of reflectivity is observed below 0.3THz (~10cm$^{-1}$). A similar result was obtained also for $(La,Eu,Sr)_2CuO_4$ [41]. These results indicate that the interlayer Josephson coupling is radically weakened as in the case of the 1/8 –phase materials, but the interlayer phase coherence survives. This is understandable as the stripe order is no more perfect for $x$>1/8 in which the incommesurability of the neutron spin signals tends to saturate or the distance between charge stripes is hardly varied despite the hole density increases [43]. It is likely that the spatial stripe correlation length would become shorter, even glassy, so that the frustration for the interlayer coupling is weakened.

2.4 Stripe phase in Nd-doped $La_{2-x}Sr_xCuO_4$ with $x$ >0.16 (overdoped regime)

Recently, it has been demonstrated by Daou *et al.* that, under magnetic fields strong enough to suppress superconductivity, the overdoped $La_{1.6-x}Nd_{0.4}Sr_xCuO_4$ undergoes a quantum transition at around $x$ = 0.23 from a metallic state involving stripe order to another metallic state, presumably a Fermi liquid [44]. On the other hand, in the phase diagram of Nd free La214, the line of the low-temperature-orthorhombic (LTO) to high-temperature-tetragonal (HTT) structural transition crosses the $T$ = 0 axis at $x$ ~ 0.22. Takagi *et al.* suggested that superconductivity would be completely suppressed above $x$ = 0.22 when homogeneous distribution of the Sr concentration were realized in a sample [45]. In fact, a rapid reduction of the superconducting volume fraction was observed when $x$ exceeds 0.22 by the μSR measurement on very high quality polycrystalline samples performed by Uemura *et al*. [46].

The $c$-axis optical reflectivity measurement was carried out to examine the interlayer phase coherence and/or bulk superconductivity in the highly Sr-doped La/Nd214 with various Nd contents. Figures 10(a) - (f) display the $c$-axis reflectivity spectra measured at several temperatures for $x$ = 0.20 and $x$ = 0.24 with $y$ = 0, 0.2, and 0.4 [47]. In addition to the dominant optical phonons above 200 cm$^{-1}$, all the spectra show a gradually rising background toward ω = 0, reflecting metallic dc (ω = 0) conductivity along the $c$-axis. The metallic $c$-axis resistivity in the highly Sr-doped region is in contrast to the insulating resistivity in the underdoped regime.

In the case of $x$ = 0.20, only for y=0.4, $\rho_c(T)$ shows a weak jump at 82 K corresponding to the LTO to LTT transition which signals the formation of the static stripe order. We see in Fig. 10 that the Josephson plasma mode shows up as a sharp edge with a pronounced dip in the $c$-axis reflectivity spectrum below $T_c$. The edge position shifts to lower frequency with increase of Nd content. The case with $y$ = 0.40 (Fig. 10(c)) is marginal. We are not sure whether or not the observed edge-like feature at $T = 5K$ (the sample shows a resistive superconducting transition at 17 K) is a Josephson plasma edge, since no remarkable change is observed between the spectrum at 20 K (above $T_c$) and that at 5 K in the frequency region above the low-frequency limit of the FIR measurement. Certainly, the situation is similar to that for Nd/Eu-La214 with $x$=0.15-0.16. The Nd content necessary for reducing the Josephson plasma frequency to below 20 cm$^{-1}$ is higher for $x = 0.20$ than that for $x = 0.15$.

The $c$-axis spectra for $x = 0.24$ are distinct from those for $x = 0.20$. Even for the Nd-free sample no sharp plasma edge appears in the spectrum even at $T = 5K$ well below the apparent superconducting $T_c \sim 15K$, and the spectrum is almost unchanged for Nd-doped samples. All the results for $x = 0.24$ indicate that neither bulk three-dimensional (3D) superconductivity nor stripe order develop in the Sr-overdoped compounds. The resistivity drop seen in these compositions may be due to fluctuation in the Sr content in a crystal. The absence of optical signature of the stripe order is consistent with the dc resistivity behavior which shows no discernible feature related to the LTO-LTT structure transition in all the samples with $x = 0.24$. The earlier neutron diffraction experiment on La$_{1.6-x}$Nd$_{0.4}$Sr$_x$CuO$_4$ with $x = 0.25$ also detected no obvious LTO-LTT transition, and the incommensurate magnetic signals at the lowest temperature was too weak to be ascribed to bulk property [48].

Here we summarize the inter-relationship between superconductivity and stripe order in Nd/La214 as follows. The present result in combination with the previous one for $x = 0.15$ [35] shows that the bulk (3D) superconductivity is more robust for higher Sr content against the formation of stripe order induced by Nd-doping in the overdoped regime. This might be ascribed to difficulty in the formation of the perfect stripe order at higher doping, as also evidenced by the results of neutron diffraction and Hall coefficient. The incommensurability of the magnetic-stripe order tends to saturate [48], and the temperature dependence of Hall coefficient does not show a radical decrease below the charge-stripe ordering temperature [49] for $x$ larger than 1/8. When $x$ exceeds the critical doping $x_c \sim 0.22$, no bulk superconductivity nor firm evidence for the formation of stripe order exist in Nd/La214. The electronic state of overdoped La214 with $x > x_c$ would be a featureless Fermi-liquid showing no superconducting instability, and the Nd-doping cannot induce the stripe order. With all these into consideration, a schematic $x$(Sr)-$y$(Nd) phase diagram ($T = 0$) for Nd/La214 is drawn in Fig. 11.

3. Experimental Indications of Nearly Static Stripe Order in Other Cuprates
3.1 La$_{2-x}$Sr$_x$CuO$_4$ and YBa$_2$Cu$_3$O$_{7-y}$ under moderate magnetic fields

Figure 1 shows how the Josephson plasma edge at $T$ well below $T_c$ evolves with doping in La214 [15]. The edge rapidly moves to higher energy with increase of $x$. This is a general trend for any cuprate system, signaling an increase of the interlayer

Josephson coupling strength with doping. However, a singular behavior is seen for La214 at $x = 0.12$ in which the position of the plasma edge is nearly the same as that for $x = 0.10$, suggestive of the manifestation of a weak interlayer decoupling effect at this particular doping close to 1/8.

The La214 system, in particular at $x = 0.12$, is widely believed to be on the verge of the stripe instability. It is likely that the stripe order is not static in this compound without LTT lattice distortions [50]. Then, the stripe fluctuations slow down upon Nd substitution and eventually become static due to the pinning by the LTT distortions.

It is interesting to investigate whether magnetic field $H$ applied parallel to the $c$-axis induces static stripe order around magnetic vortices in the superconducting state of La214. Actually, a low-energy shift of the Josephson plasma edge is observed at moderate fields much lower than the pair-breaking field $H_{c2}^{ab}$ [51, 52]. As indicated in the inset of Fig. 12, the decrease of the $c$-axis superconducting condensate $\rho_{s,c}$ (that is proportional to the square of Josephson plasma frequency or $\omega_{ps}^2$) with $H$ is remarkably fast for the underdoped La214 ($x = 0.10$ and 0.12), and the plasma edge fades out at a moderate magnetic field $H // c$ (~ 8 Tesla) [51], in quite a similar manner to that already seen in the case of Nd doping into La214.

The local pair-breaking in the vicinity of each vortex within a layer does not give an immediate answer for the suppression of the Josephson plasma frequency. In the $H$-$T$ phase diagram ($H // c$) of HTSC, there is a wide "vortex liquid" region where vortices are no longer straight lines along the c-axis and cannot form the Abrikosov lattice because of the two-dimensional nature of the electronic state. Schafgan et al. carefully examined the spectral change from the viewpoint of this vortex motion [51]. For highly doped La214 with $x$=0.15 and 0.17, they found that the decrease of $\rho_{s,c}$ due to magnetic field can be understood within the vortex wandering model [52] in which the superconducting phase coherence along the $c$-axis is disturbed by vortices displacement between neighboring $CuO_2$ layers (see the insets of Fig. 12).

In contrast, for the underdoped samples, an additional mechanism is necessary to explain the accelerated decrease of $\rho_{s,c}$ with field. Schafgan et al. ascribe this to the stabilization of fluctuating magnetism by magnetic field, as was reported in the neutron experiment [53] which indicates that the stabilized magnetic order extends to a length scale much longer than the in-plane superconducting coherence length ($\xi_0$ ~ 2–3 nm). It is highly plausible that the induced magnetism is associated with that in the anti-ferromagnetic spin-stripe domains between charge stripes, since otherwise it would be difficult to explain such a strong suppression of interlayer coupling by moderate magnetic fields.

In this regard, a similar rapid decrease of $\rho_{s,c}$ observed for underdoped $YBa_2Cu_3O_{6.6}$ ($T_c \sim 60$ K) [54] is suggestive of a magnetic-field induced stripe-like order (see Fig. 13). Y123 also exhibits incommensurate spin response over a wide doping range, and in particular, a nematic spin order with broken rotational symmetry has recently been discovered in the underdoped pseudogap region, reminiscent of the stripe order in La-based 214 compounds[44]. The reduction of Josephson plasma frequency with magnetic field is shown in Fig. 14 for La214 and Y123 with similar doping levels ($x \sim$ 0.12 and 0.15). For both systems, the compounds with $x \sim 0.12$ show a stronger suppression of the interlayer Josephson coupling with magnetic field, suggestive of similar dephasing effect associated with similar order.

3.2 Transverse Josephson plasma in La214 with $x$ near 1/8

The presence of a transverse Josephson plasma in La214 was originally proposed by van der Marel and Tsvetkov [23]. They attributed the anomalous conductivity peak just above the reflectivity plasma edge and/or the asymmetric broadening of plasmon peak in the loss function to the modulation of the Josephson coupling strength along the $c$-axis. As an origin of Josephson coupling modulation, a random substitution of Sr for La was considered, but it is not clear how the random distribution of Sr dopants gives rise to a periodic modulation of the interlayer Josephson coupling strength.

The relation of this phenomenon to the stripe order was discussed by Dordevic *et al.* [55] The reflectivity dip associated with Josephson plasma edge becomes remarkably shallow for $x$ near 0.125, despite a sharp drop in reflectivity from 100 % level. This anomalous spectral shape in reflectivity was attributed to the presence of an unusual absorption peak just above the edge, which makes the peak in the loss function asymmetrically broadened (Fig. 15). The fact that the anomalous spectral feature becomes most pronounced at $x = 1/8$ suggests that the observed 'anomaly' may be related to (dynamical or short-range) stripe order.

Another support for the presence of transverse Josephson modes is obtained by the recent measurement for $x = 0.13$. using a reflection-type THz-TDS [56] as well as a coherent-source spectrometer in a submilli- and millimeter wavelength region [57]. While the absolute value of reflectivity and weak spectral features are not accurately determined, prominent spectral features, such as the Josephson plasma edge and optical phonon bands, and their temperature dependences can be reliably caught by the THz-TDS. As shown in Fig. 16 [56], the reflectivity spectra show a bump (at ~ 1.4 THz) below $T_c$ in higher frequency side of the Josephson plasma edge, both shifting towards higher frequency with lowering temperature. This cannot be reproduced by a conventional two-fluid model, but the spectral feature resembles the spectra shown in Fig. 4 for T*-214. An advantage of THz-TDS is that it can directly determine complex conductivity (or dielectric function) without Kramers-Kronig analysis. The result (right panel of Fig. 16) shows that a conductivity peak shifts with temperature below $T_c$, the behavior expected for a transverse plasma mode.

In order to have a transverse plasma mode in a single-layer cuprate, a periodic modulation of the Josephson coupling strength along the c-axis has to be realized as in the case of T*214. A random distribution of Sr cannot create a well-defined mode. A possible origin may be a sequential stack of 'inequivalent' layers such as that in the ideal stripe order, schematically depicted in Fig. 8. The ideal stripe order, two pairs of the striped double layers, in which the charge stripes in one layer are orthogonal to those in the next, completely frustrates interlayer Josephson coupling. As we have seen in the preceding sections, the interlayer dephasing effect dominates in the static stripe ordered state of La/Nd214, even if the order is not perfect. To the contrary, if the Josephson coupling between neighboring layers is intact, but the interlayer couplings are weakly modulated by fluctuating and/or short-range stripe order, which seems to be the case with La214 with $x$ near 1/8, then the Josephson coupling strength might be modulated along the c-axis, giving rise to a transverse mode like that in T*214.

4. Effects of Stripe Order on the In-Plane Optical Spectra
4.1 In-plane normal-state spectrum in the stripe ordered phase

Compared to the c-axis optical studies, the researches of the in-plane optical spectra for the stripe materials are limited. We first show the in-plane spectra of the nickelate $(La,Sr)_2NiO_4$ which is also a doped Mott insulator in Fig. 17. The spectrum of $La_{1.67}Sr_{0.33}NiO_4$ which undergoes a charge-stripe ordering transition at $T = 240$ K (spins order at lower temperature, ~ 190K) and is insulating at low temperatures [8] shows a typical charge ordering effect. As temperature decreases from 300 K, after a gradual suppression of low-ω conductivity probably due to fluctuating charge order, a charge-order gap associated with the long-range stripe order starts to open below 240 K and reaches the energy of 0.26 eV at 10 K. This spectral behavior is typical of that observed above and below the transition temperature of some kind of density waves.

A spectral change observed for $La_{2-x}Ba_xCuO_4$ with $x$ = 1/8 [36,58] is less dramatic. Figure 18 displays temperature dependence of the in-plane optical conductivity spectrum for this compound. From 300 to 60 K, the low-ω conductivity increases as a result of narrowing of the Drude component (Fig. 18(a)). In contrast, below the stripe order temperature (the charge order at 54K and the spin order at 42K), the low-ω conductivity (< 300 cm$^{-1}$) is suppressed, which is partly transferred to higher frequencies. The nearly $T$-independent resistivity (see Fig. 2(b) in ref.[36]) is thought to be a consequence of a decrease in the Drude weight with its width continuing to be narrowed. Coexistence of a Drude component and a charge-gap excitation is compatible with the picture of conducting stripes.

The in-plane spectrum in the stripe ordered state of La/Nd214 looks fairly different. Firstly, we cannot see any impact of the LTO-LTT phase transition at $T_d$ ~70 K on the optical spectra, despite that the resistivity shows a jump and a static charge stripe order is supposed to set in at this temperature [59,60]. Instead of gap opening at $T_d$, the low ω–conductivity is gradually suppressed from much higher temperatures, and a broad peak develops centered around 100 cm$^{-1}$ in the Nd-doped La214 with Sr-content $x$ = 0.10 and 0.12 [59] (see Fig. 19(a)). This seems a signature of tendency toward carrier localization. The suppressed low-ω spectral weight is transferred to higher energy region above 1 eV. This energy scale of spectral weigh transfer is larger than the case of La$_{1.875}$Ba$_{0.125}$CuO$_4$ [58].

Figure 19(b) is the optical conductivity spectra of 60 % Nd-doped La214 with $x$ = 0.125 obtained by another research group [60]. Although they did not observe a conductivity suppression due to Nd-doping, they found only a peak around 100 cm$^{-1}$ growing gradually with lowering temperature, similar to that in Fig.19(a). This peak was ascribed to a tendency for charge localization in the stripe ordered state.

The difference between LBCO and La/Nd214 may be attributed to the effect of disorder and/or to the difference in the amplitude of both the electronic and lattice modulations associated with stripe order. About 6% of the La sites are replaced by Ba in LBCO, whereas the fraction of the replacement is as large as 26% in La/Nd214 by Nd and Sr substitution. The crystalline lattice of La/Nd214 is, therefore, more disordered. According to the neutron and x-ray studies the spatial correlation length of the stripe order is by a factor of two longer, and the modulation amplitude is also significantly larger in LBCO [58]. As superconductivity is appreciably affected by the disorder at the La sites, the charge dynamics in the stripe ordered state may have some influence from such disorder.

4.2 Effect of superconducting stripes on the in-plane spectrum

The in-plane optical spectrum of La$_{2-x}$Ba$_x$CuO$_4$ with $x$ = 0.125 was measured at $T$ = 6 K below $T_{BKT}$ ~16 K but above bulk $T_c$ ~2.4 K by Homes et al. [36, 58]. They observed a noticeable decrease of conductivity below ~300 cm$^{-1}$ (Fig.18(b)), but the lost spectral weight is not transferred to a delta function at ω = 0 but to higher frequency region.

So far, there is no published data taken at temperature below $T_c$ for this compound.

On the other hand, for $La_{1.45}Nd_{0.4}Sr_{0.15}CuO_4$ the in-plane spectrum shows superconducting response below $T_c$ = 12 K [59]. One can see a small but finite missing area in the conductivity spectrum shown in Fig. 20 but the spectrum at temperature well below $T_c$ is dominated by a huge residual conductivity in the low-$\omega$ region, suggestive of the presence of a large amount of uncondensed electrons. In such a situation, it is difficult to accurately estimate a condensate weight from the missing area.

An alternative way to estimate a condensate is to measure surface impedance in the microwave region. The microwave impedance was measured on the same sample, and the in-plane penetration depth $\lambda_{ab}$ at $T$ = 0 was estimated as 0.8–1.0 μm [59]. This is by a factor of 3–4 longer than $\lambda_{ab}$ of the Nd-free sample, indicating that the in-plane condensate weight is radically reduced by the formation of stripe order. In this respect, it is important for the notion of 'superconducting stripes' to know if superconductivity and stripe order coexist in the same volume of the stripe ordered material. The earlier μSR study of superconductivity and magnetism on the $La_{1.85-y}Eu_ySr_{0.15}CuO_4$ system [62] suggested that superconductivity and magnetism (associated with the stripe order) occur in separate volumes of the system, but the length scale of such phase separation remains to be determined.

5. Summary

The most remarkable effect of the stripe order on superconductivity is the interlayer decoupling. The HTSC is generally an array of the superconducting $CuO_2$ layers stacked along the $c$-axis. A bulk superconducting state with interlayer phase coherence is achieved by Josephson coupling which is evidenced by an appearance of a collective Josephson plasma mode in the $c$-axis optical spectrum. A rapid suppression of the interlayer phase coherence as the stripe order develops is demonstrated by a reduction of the Josephson plasma frequency with increase of the Nd content in La/Nd214.

In real materials showing the stripe order, a complete interlayer decoupling is difficult to be achieved due to an inhomogeneous distribution of dopant Ba or Sr atoms and/or disorder introduced by the dopant atoms themselves and by the incorporation of Nd atoms to stabilize the stripe order. A Josephson plasma edge, albeit not a perfect form, is observable in the $c$-axis reflectivity spectra even for compounds with $x$ near 1/8, when measurement is extended down to sub-THz region. For La/Nd214 the stripe order is stabilized over a wide range of doping, and the suppression of condensate weight due to the stripe order is more dramatic in the $c$-direction than in the $ab$-direction, pointing to two-dimensional superconductivity.

On the other hand, in the highly doped ($x > 1/8$) compounds the $c$-axis phase coherence is more robust, probably because the correlation length of stripes becomes shorter. In the overdoped region, specifically for $x > 0.22$, no Josephson plasma mode is identified even for Nd-free La214, and no evidence for the stripe order is obtained. This indicates that the overdoped La214 cannot sustain both superconductivity and stripe order.

Nearly ideal stripe order seems to be formed in LBCO with $x = 1/8$ and in La/Nd214 with $x \sim 0.12$. as the bulk $T_c$ is suppressed to 2–4 K. Several years after the optical study which suggested interlayer decoupling in the stripe phase[35], the 2D superconductivity was independently found by the other experimental techniques in $La_{1.875}Ba_{0.125}CuO_4$. The development of the superconducting order within a $CuO_2$ layer at temperatures $T_{BKT}$ well above $T_c$ was suggested by the transport [37] and ARPES measurements [63]. In fact, no plasma mode is observed below $T_{BKT}$, although no firm evidence for superconductivity in the layer has so far been obtained in the in-plane optical spectrum.

In the other cuprates, Nd-free La214 and Y123, there is no evidence of static stripe order, and the $c$-axis phase coherence is established below very high $T_c$. Nevertheless, an anomalously rapid reduction of the Josephson plasma frequency is observed for underdoped La214 and Y123, when moderate magnetic fields are applied perpendicular to the $CuO_2$ layers. The interlayer decoupling due to vortex wandering is normally weak. So, the tendency for strong interlayer decoupling might be ascribed to an effect similar to the decoupling by stripe order.

Acknowledgement
The authors thank K. Tanaka at Osaka University and K. Kojima at KEK, Japan for providing their unpublished data and the fruitful discussions.

Figure Captions

Fig.1 Reflectivity spectra of $La_{2-x}Sr_xCuO_4$ for $E // c$. (From ref.[15])

Fig.2 Schematic picture of Josephson coupled $CuO_2$ layers with coupling strength alternately modulated along the $c$-axis (Cross section of HTSC along the $c$-axis).

Fig.3 Conductivity spectra of $La_{2-x}Sr_xCuO_4$ for $E // c$ at temperatures above (40K or 35K) and below $T_c$ (25K and 8K). (From ref. [20])

Fig.4 The $c$-axis reflectivity spectra of $SmLa_{1-x}Sr_xCuO_4$ with $x = 0.15$ ($T_c = 30$ K) and 0.2 ($T_c = 25$ K; inset). The weak peak developing below $T_c$ just above the plasma edge frequency originates from the transverse Josephson plasma mode. (From ref.[25])

Fig.5 Structural and electronic phase diagram for $La_{1.85-y}Nd_ySr_{0.15}CuO_4$.
The data for the structural phase transition from LTT to LTO are taken from ref.[33]. The $T_c$ values are from ref.[35]

Fig.6 The $c$-axis reflectivity spectra of $La_{1.85-y}Nd_ySr_{0.15}CuO_4$ at 8 K for various Nd contents. (From ref.[35])

Fig.7 Nd content dependence of $T_d$, $T_c$ and $\omega_{ps}^2$ for $La_{1.85-y}Nd_ySr_{0.15}CuO_4$. Solid curves are guides for the eyes. (From ref.[35])

Fig.8 Schematic picture of the charge/spin stripes

Fig.9 The $c$-axis reflectivity spectra of $La_{1.84-y}Nd_ySr_{0.16}CuO_4$ with $y=0.3$ ($T_c = 20$ K) and 0.4 ($T_c = 12$ K) measured by a time domain spectroscopy in the THz region. (From ref.[42])

Fig.10 The $c$-axis reflectivity spectra of $La_{1.8-y}Nd_ySr_{0.2}CuO_4$ with $y=0$ (a), 0.2 (b) and 0.4 (c), and $La_{1.76-y}Nd_ySr_{0.24}CuO_4$ with $y=0$ (d), 0.2 (e) and 0.4 (f). (From ref.[47])

Fig.11 Schematic picture of the $x$-$y$ phase diagram for La/Nd214.
HTT: <u>H</u>igh <u>T</u>emperature <u>T</u>etragonal, LTT: <u>L</u>ow <u>T</u>emperature <u>T</u>etragonal, LTO:

Low Temperature Orthorhombic, LTLO: Low Temperature Less-orthorhombic

Fig.12 The loss function for the $c$-axis optical spectra of $La_{2-x}Sr_xCuO_4$ at 8 K in various magnetic fields $H // c$. The insets show the $H$-dependence of normalized Josephson plasma frequency squared. (From ref.[51])

Fig.13 The c-axis reflectivity spectra of underdoped $YBa_2Cu_3O_y$ with $T_c$ = 59 K. in various magnetic fields applied parallel to the $c$-axis (From ref.[54])

Fig.14 Magnetic field $H(//c)$ dependence of the squared Josephson plasma frequency for lightly ($T_c$ = 76 K) and heavily ($T_c$ = 59 K) underdoped Y123. The data of La214 are also plotted for comparison. (From ref.[54])

Fig.15 Conductivity spectra and loss function of $La_{2-x}Sr_xCuO_4$ for $E // c$. (From ref.[55])

Fig.16 The $c$-axis reflectivity (left) and conductivity (right) spectra of $La_{1.87}Sr_{0.13}CuO_4$ measured by THz time domain spectroscopy. $T_c$ = 36 K. (From ref.[56])

Fig.17 Temperature evolution of the in-plane optical conductivity of $La_{1.67}Sr_{0.33}NiO_4$ showing a charge ordering at 240 K. (From ref.[8])

Fig.18 Temperature dependence of the in-plane optical conductivity spectrum of $La_{1.875}Ba_{0.125}CuO_4$. At 6K($<T_{BKT}$) the Drude spectrum becomes too narrow to be observed within a measurement wave number range. The insets show the in-plane reflectivity spectra. (From ref.[36])

Fig.19 The in-plane conductivity spectra of $La_{2-x-y}Nd_ySr_xCuO_4$ with $x$ = 0.12, $y$ = 0.4 (a), and $x$ = 0.125, $y$ = 0.6 (b). (From (a)ref.[59] and (b)ref.[60])

Fig.20 Temperature dependence of the in-plane optical conductivity for $La_{1.6-x}Nd_{0.4}Sr_xCuO_4$ with $x$ = 0.15. (From ref.[59])

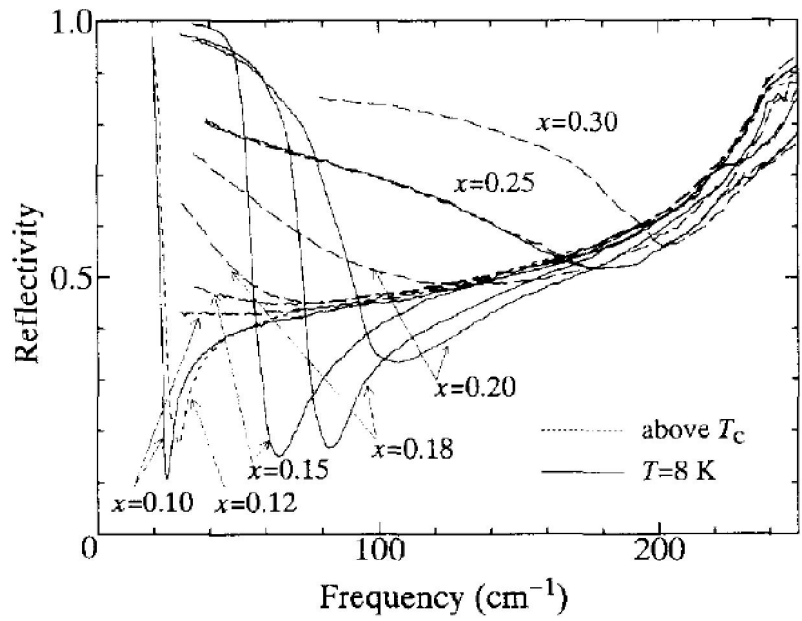

Fig.1   Reflectivity spectra of $La_{2-x}Sr_xCuO_4$ for $E // c$.   (From ref.[15])

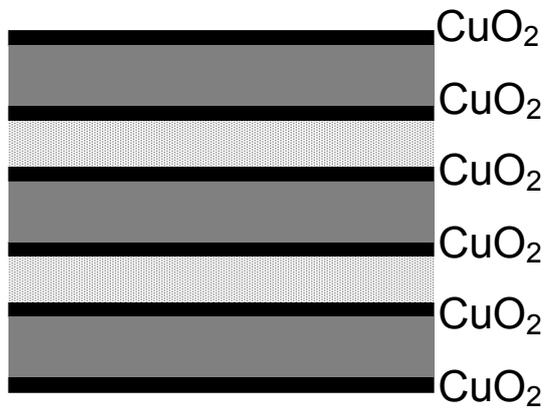

Fig.2 Schematic picture of Josephson coupled $CuO_2$ layers with coupling strength alternately modulated along the $c$-axis (Cross section of HTSC along the $c$-axis).

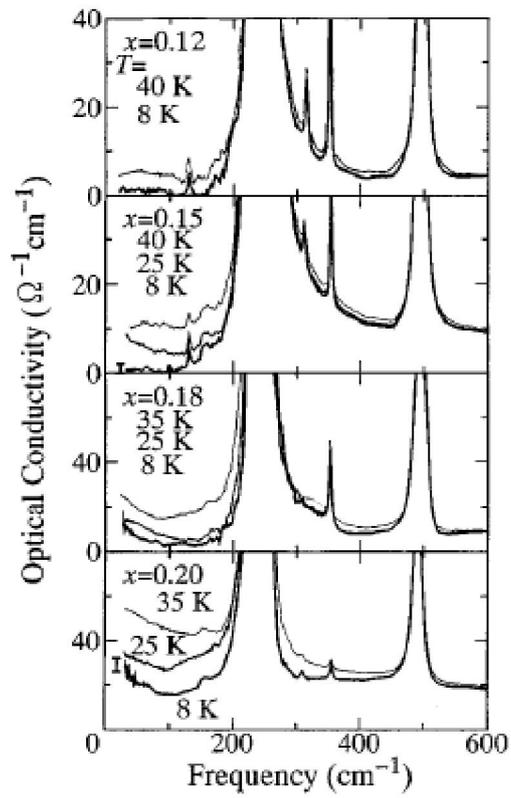

Fig.3 Conductivity spectra of $La_{2-x}Sr_xCuO_4$ for $E // c$ at temperatures above (40K or 35K) and below $T_c$ (25K and 8K). (From ref. [20])

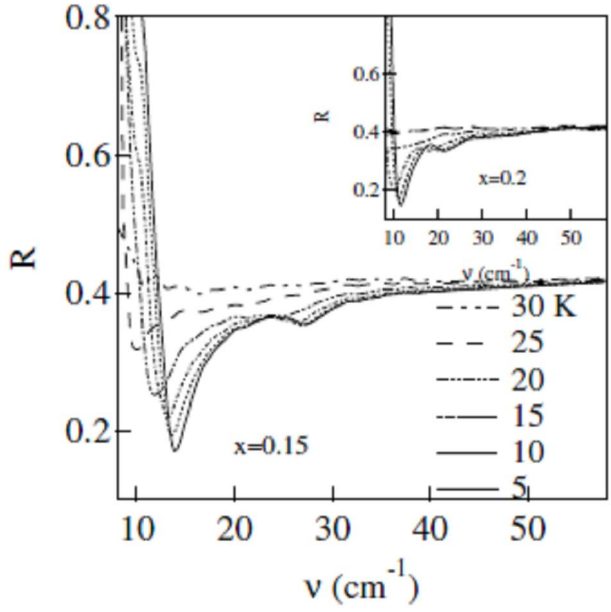

Fig.4  The $c$-axis reflectivity spectra of SmLa$_{1-x}$Sr$_x$CuO$_4$ with $x$ = 0.15 ($T_c$ = 30 K) and 0.2 ($T_c$ = 25 K; inset). The weak peak developing below $T_c$ just above the plasma edge frequency originates from the transverse Josephson plasma mode.   (From ref.[25])

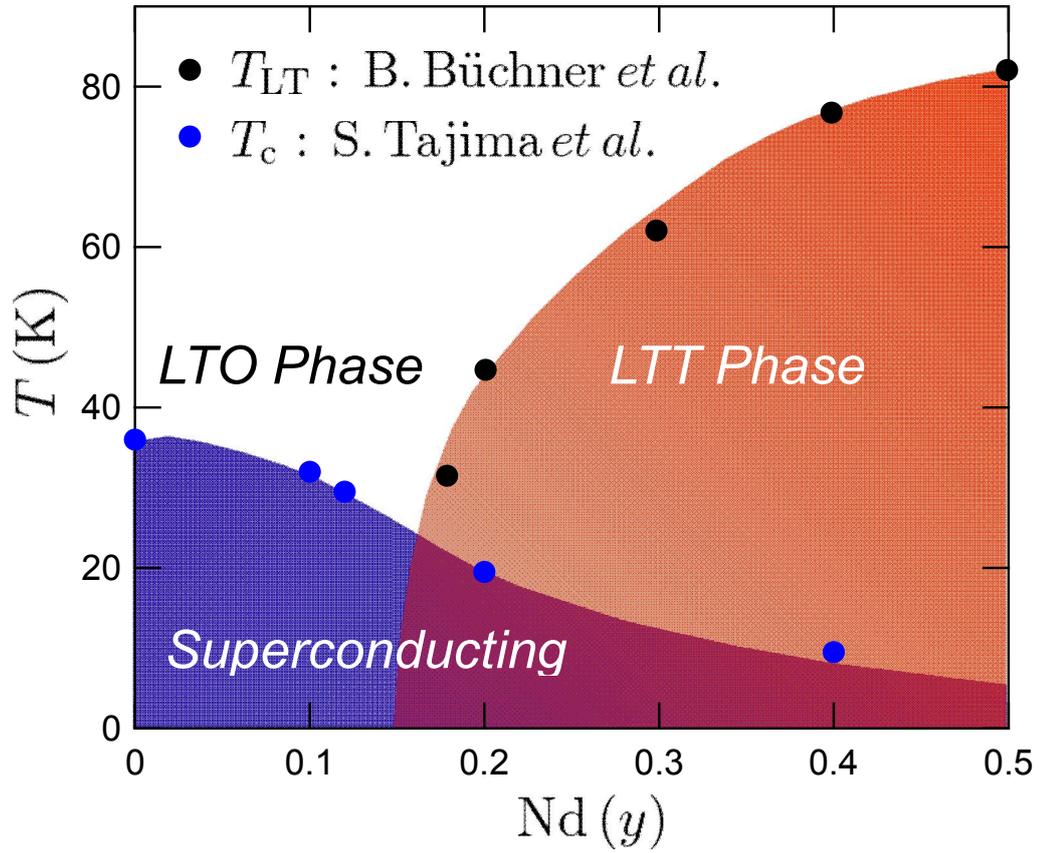

Fig.5 Structural and electronic phase diagram for $La_{1.85-y}Nd_ySr_{0.15}CuO_4$.
The data for the structural phase transition from LTT to LTO are taken from ref.[33].
The $T_c$ values are from ref.[35]

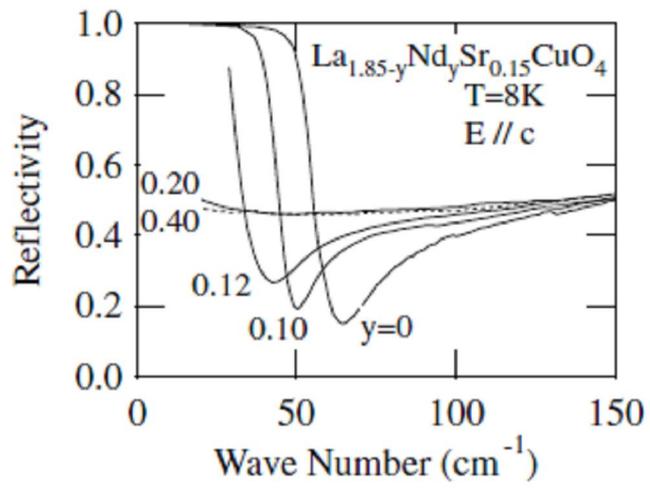

Fig.6 The $c$-axis reflectivity spectra of $La_{1.85-y}Nd_ySr_{0.15}CuO_4$ at 8 K for various Nd contents. (From ref.[35])

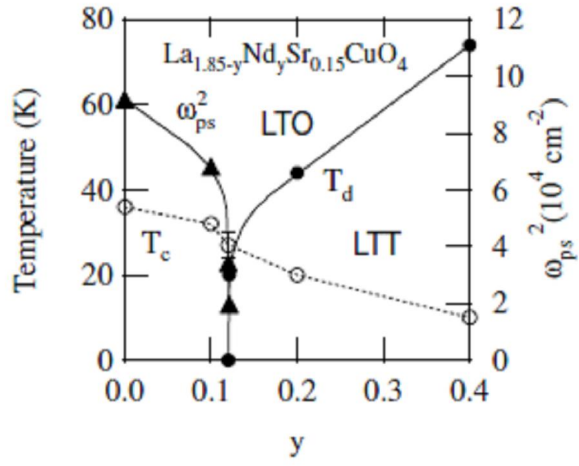

Fig.7 Nd content dependence of $T_d$, $T_c$ and $\omega_{ps}^2$ for $La_{1.85-y}Nd_ySr_{0.15}CuO_4$. Solid curves are guides for the eyes. (From ref.[35])

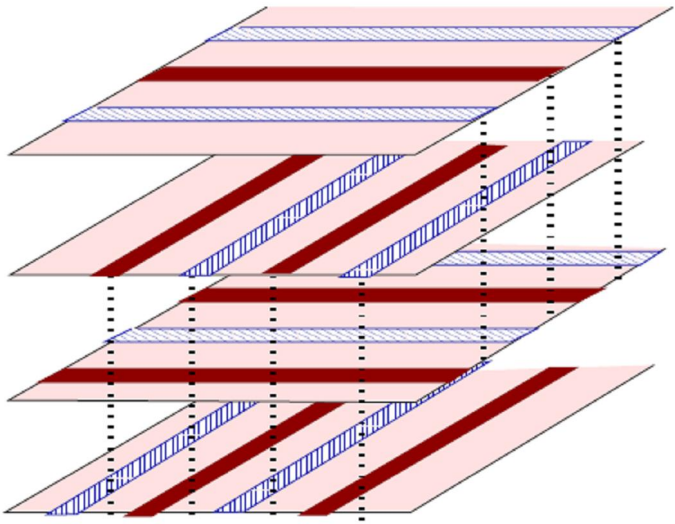

Fig.8 Schematic picture of the charge/spin stripes

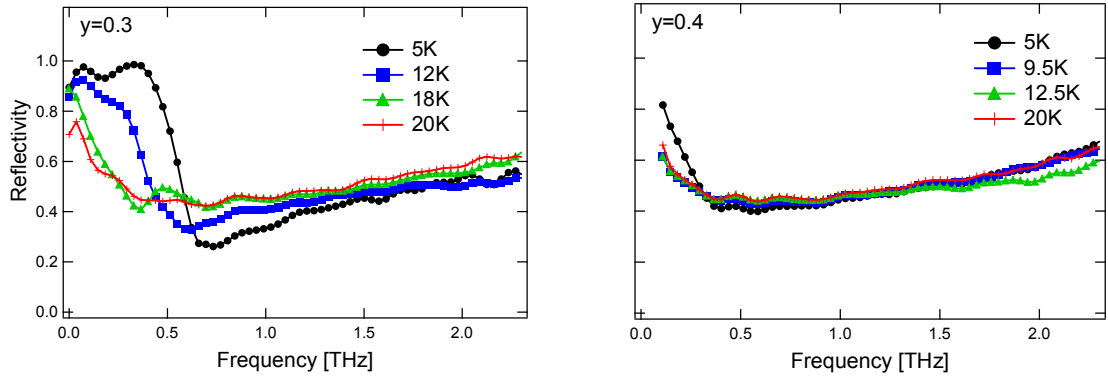

Fig.9 The $c$-axis reflectivity spectra of $La_{1.84-y}Nd_ySr_{0.16}CuO_4$ with $y$=0.3 ($T_c$ = 20 K) and 0.4 ($T_c$ = 12 K) measured by a time domain spectroscopy in the THz region. (From ref.[42])

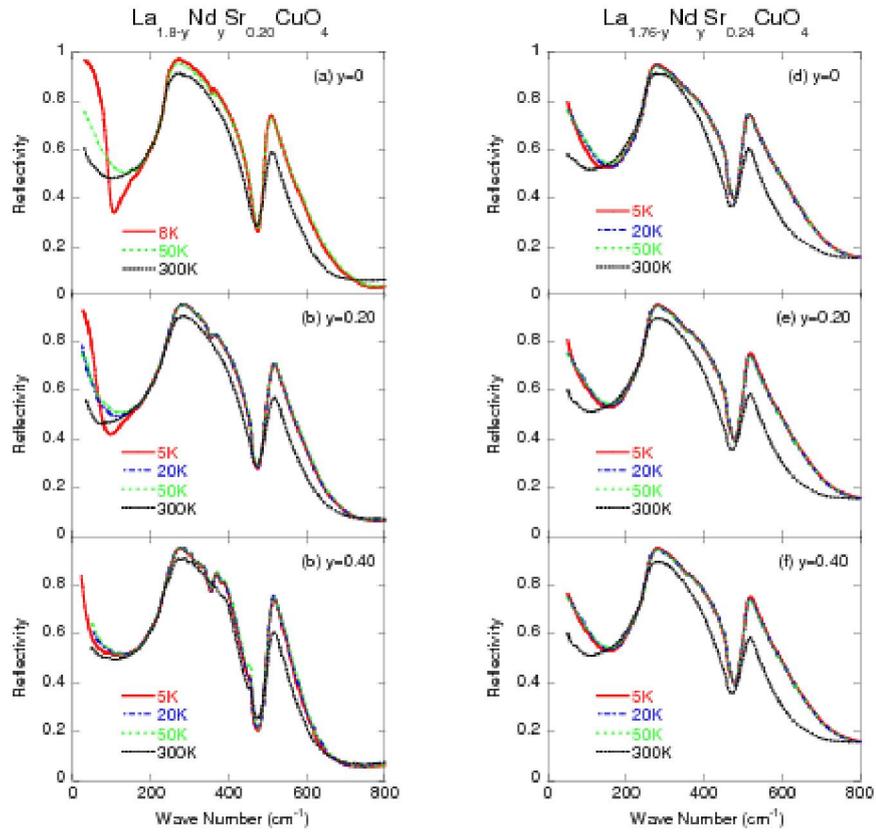

Fig.10 The $c$-axis reflectivity spectra of $La_{1.8-y}Nd_ySr_{0.2}CuO_4$ with (a) $y=0$ ($T_c$=30K), (b) y=0.2 ($T_c$=23K) and (c) y=0.4 ($T_c$=17K), and $La_{1.76-y}Nd_ySr_{0.24}CuO_4$ with (d) $y=0$ ($T_c$=15K), (e) y=0.2 ($T_c$=12K) and (f) y=0.4 ($T_c$=7K).   (From ref.[47])

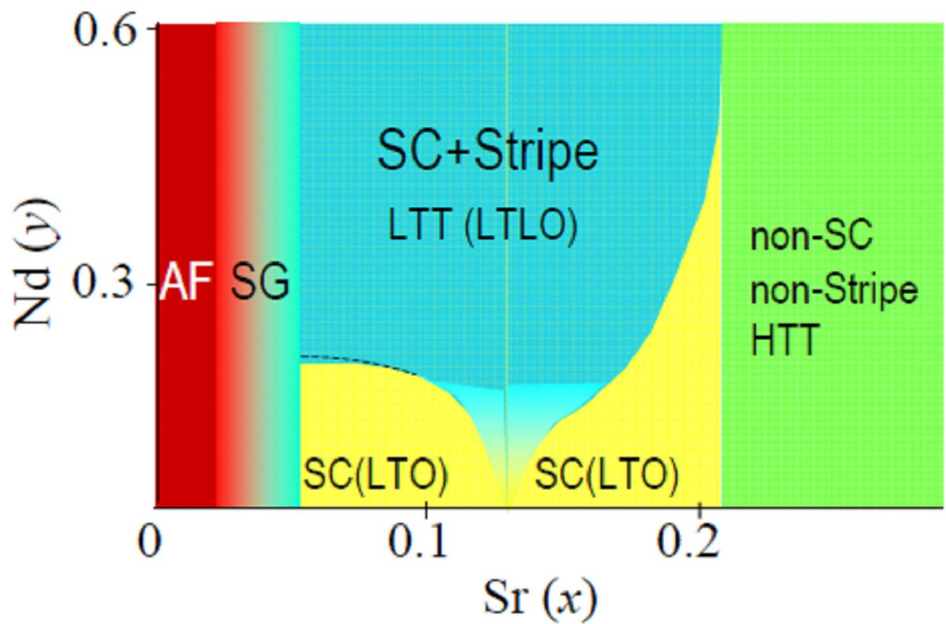

Fig.11 Schematic picture of the *x-y* phase diagram for La/Nd214.

HTT: High Temperature Tetragonal, LTT: Low Temperature Tetragonal, LTO: Low Temperature Orthorhombic, LTLO: Low Temperature Less-orthorhombic

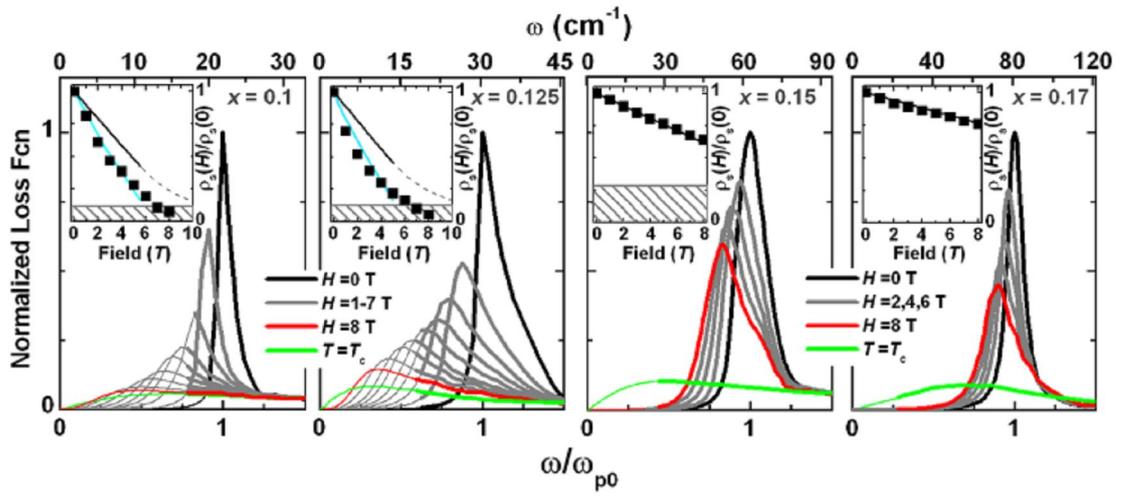

Fig.12 The loss function for the $c$-axis optical spectra of $La_{2-x}Sr_xCuO_4$ at 8 K in various magnetic fields $H // c$. The insets show the $H$-dependence of normalized Josephson plasma frequency squared. (From ref.[51])

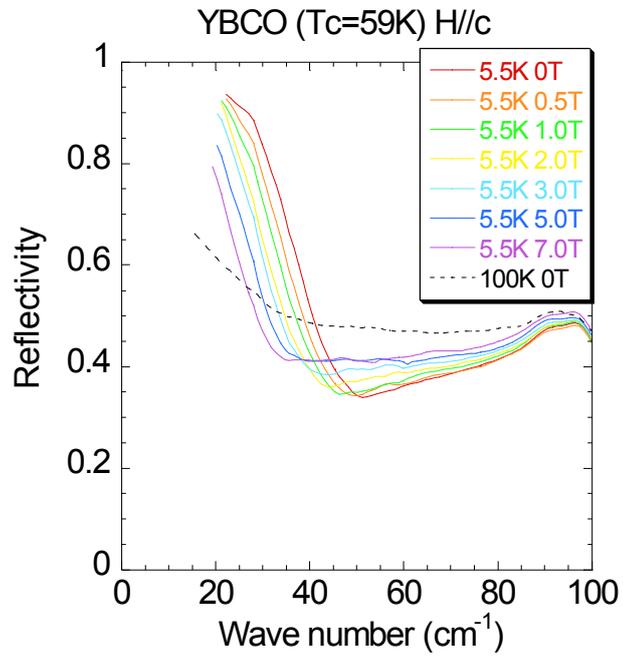

Fig.13 The c-axis reflectivity spectra of underdoped $YBa_2Cu_3O_y$ with $T_c$ = 59 K. in various magnetic fields applied parallel to the $c$-axis (From ref.[54])

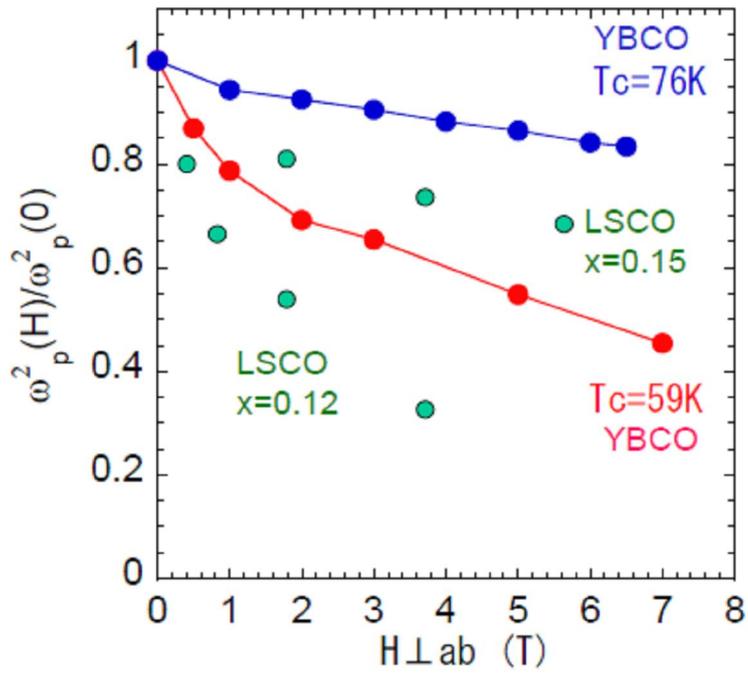

Fig.14 Magnetic field $H(//c)$ dependence of the squared Josephson plasma frequency for lightly ($T_c$ = 76 K) and heavily ($T_c$ = 59 K) underdoped Y123. The data of La214 are also plotted for comparison. (From ref.[54])

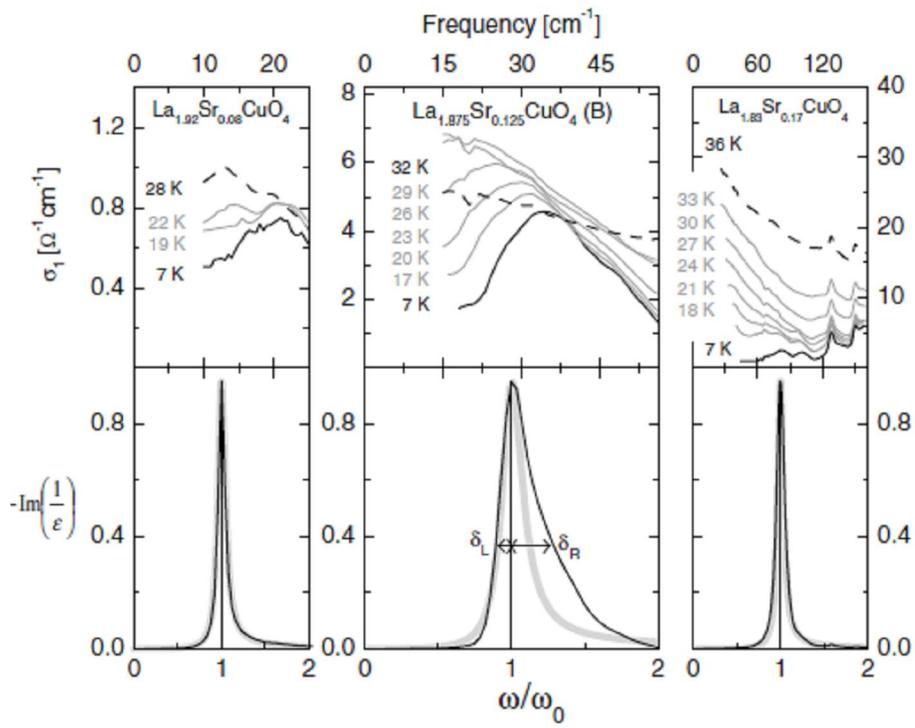

Fig.15 Conductivity spectra and loss function of $La_{2-x}Sr_xCuO_4$ for $E \parallel c$. (From ref.[55])

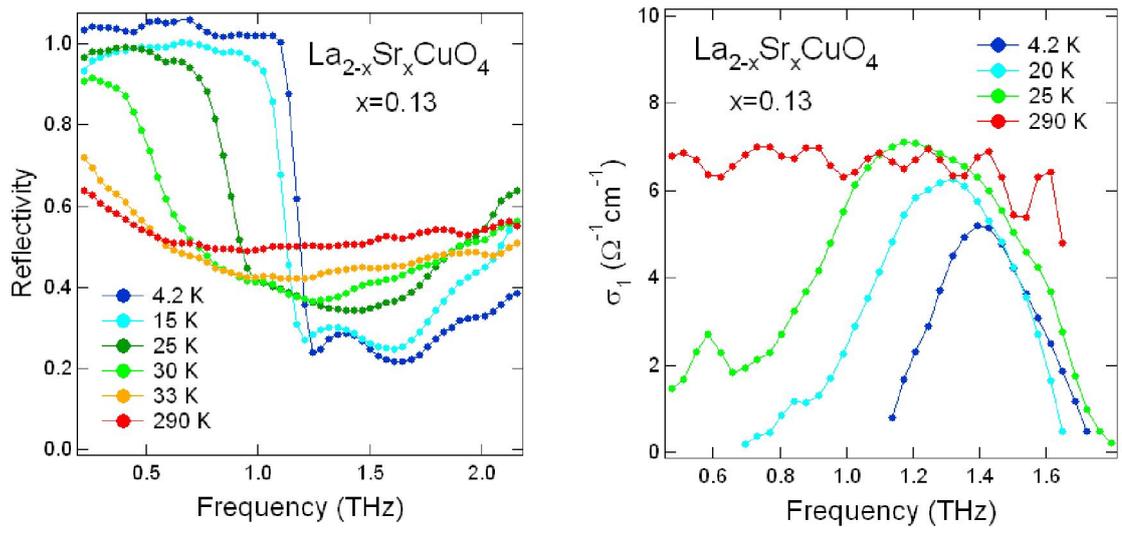

Fig.16 The *c*-axis reflectivity (left) and conductivity (right) spectra of La$_{1.87}$Sr$_{0.13}$CuO$_4$ measured by THz time domain spectroscopy. $T_c$ = 36 K. (From ref.[56])

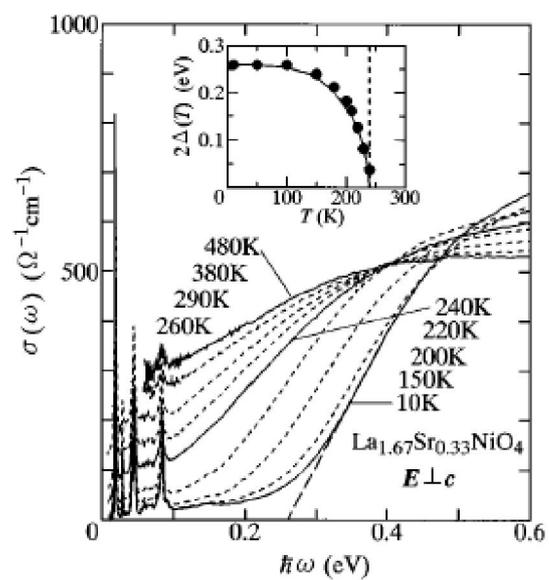

Fig.17 Temperature evolution of the in-plane optical conductivity of $La_{1.67}Sr_{0.33}NiO_4$ showing a charge ordering at 240 K. (From ref.[8])

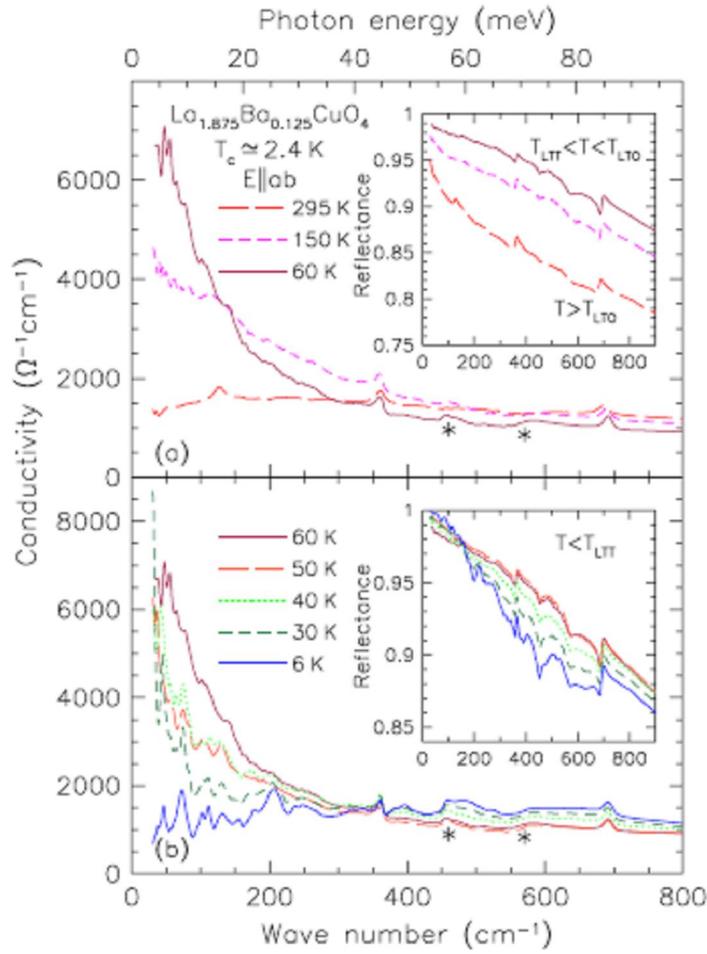

Fig.18 Temperature dependence of the in-plane optical conductivity spectrum of $La_{1.875}Ba_{0.125}CuO_4$. At 6K($< T_{BKT}$) the Drude spectrum becomes too narrow to be observed within a measurement wave number range. The insets show the in-plane reflectivity spectra. (From ref.[36])

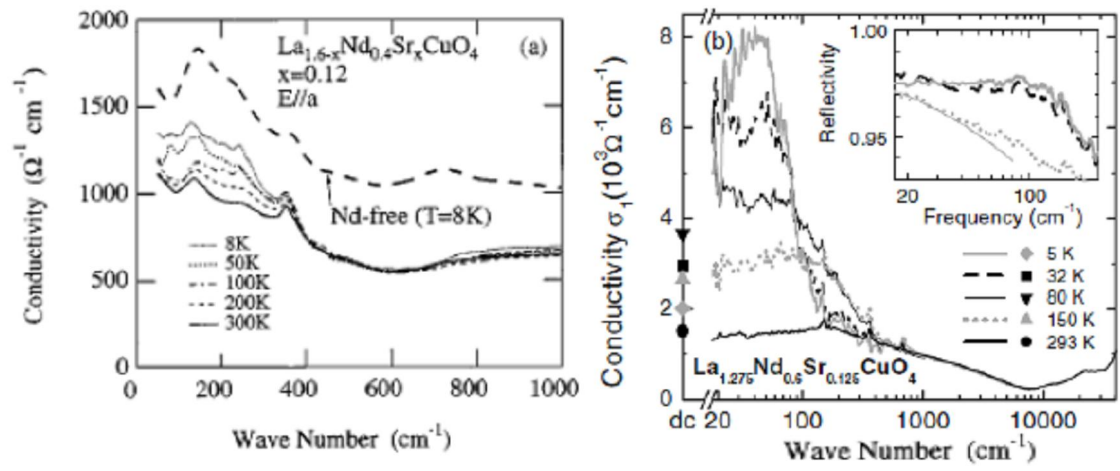

Fig.19 The in-plane conductivity spectra of $La_{2-x-y}Nd_ySr_xCuO_4$ with $x = 0.12$, $y = 0.4$ (a), and $x = 0.125$, $y = 0.6$ (b). (From (a)ref.[59] and (b)ref.[60])

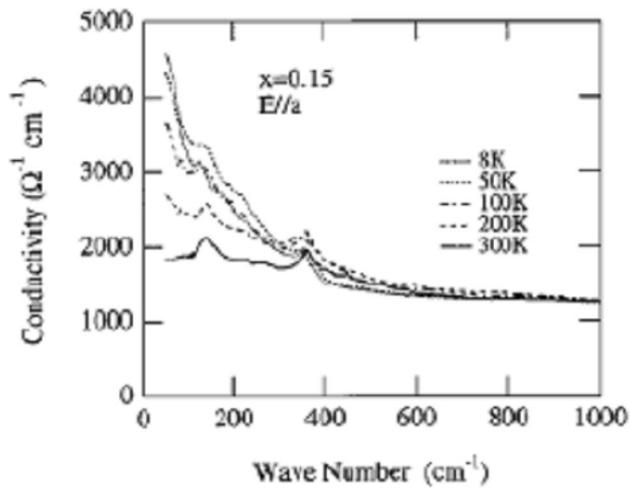

Fig.20 Temperature dependence of the in-plane optical conductivity for $La_{1.6-x}Nd_{0.4}Sr_xCuO_4$ with $x = 0.15$. (From ref.[59])